\documentstyle{article}

\title{Order   $m \alpha^8$ contributions to the decay rate
of Orthopositronium }
\author{Patrick Labelle and G. Peter Lepage
  \\ \small Newman Laboratory
of Nuclear Studies, Cornell University, Ithaca, NY 14853   \and
Ulrika Magnea\thanks{Present
 address: Dipartimento di fisica teorica dell'Universita'
 di Torino, Via P. Giuria 1,
10125 Torino, Italy.} \\ \small State University of New York
 at Stony Brook, Stony
Brook, NY 11794}

\begin{document}
\maketitle
\begin{abstract}
We discuss how  contributions to the
 order ${\cal O} (m \alpha^8)$ orthopositronium decay rate
can be separated into two categories, one  due to relativistic momenta
and calculable in terms of QED scattering amplitudes,
 the other due to low momenta and calculable in the simpler framework of
a low-energy effective theory (NRQED). We report new results for all
low-momentum contributions, and give a formula relating the
remaining contributions to conventional (on-shell) QED
scattering amplitudes.
\end{abstract}

Despite its many successes, quantum electrodynamics has yet to account
fully for the decay rate of orthopositronium. The theoretical expression
for this rate is \cite{Lepage,Karsh}
\begin{eqnarray}
 \lambda_{th}~=~{ \alpha^6 m c^2 \over \hbar}{(2 \pi^2- 18) \over 9 \pi}
\biggl[1 -10.282(3) ({\alpha \over \pi}) \nonumber  \\   +
 {1 \over 3} \alpha^2~\ln \alpha +B \,({\alpha
\over \pi})^2 -{3 \over 2 \pi}
  \,\alpha^3 \,( \ln \alpha )^2\ldots\biggr]   \label{expansion}
\end{eqnarray}

\noindent where the coefficient
  $B$   has not yet been computed.  The measured rate is
$ \lambda_{exp}= 7.0482(16)  \mu {\rm s}^{-1}$ \cite{exp}.
 The difference between this value and the known part of $\lambda_{th}$,
\begin{eqnarray}
\lambda_{exp} - \lambda_{th} (B=0)~=~ 99(16) \times 10^{-4} \mu {\rm s}^{-1}
\end{eqnarray}
is surprisingly large: the coefficient $B$ in $\lambda_{th}$ would have to be
about 250
  to bring theory and experiment into agreement. Such a large
coefficient would be unusual, but is by no means impossible, particularly
given the big ${\cal O } (\alpha/ \pi)$ correction. A complete calculation of
$B$ is
essential before any realistic assessment of the situation is possible.

A calculation of these coefficients using traditional bound-state methods
is very complicated. This is because each term in a traditional expansion
has contributions from both nonrelativistic and relativistic momenta. This
means that approximations only valid for small $p$ or only for large $p$ cannot
be readily employed to simplify the analysis.
In this paper, we outline a new and simpler procedure for computing
$B$. Our analysis is based upon a rigorous nonrelativistic
reformulation of QED called nonrelativistic quantum electrodynamics (NRQED)
\cite{NRQED}.
Using this effective field theory, we are able to separate the ${\cal O}
(\alpha^2)$ corrections into three parts. Two of these involve
 soft, nonrelativistic
momenta and probe the bound-state nature of the system. The other
part involves hard, relativistic momenta and is therefore largely
insensitive to the details of binding. We have calculated the
nonrelativistic contribution and present the results here. We also
show how to extract the relativistic contribution from a
calculation of ordinary scattering amplitudes;
no bound-state physics is required  in this part  of the calculation.

Our positronium results also have  implications for quarkonium decays. These
will be discussed in another paper \cite{quark}.

The lowest-order decay
rate of orthopositronium (\mbox{O-Ps}, $n=J=S=1$) is given by
(we now use natural units, with $c= \hbar =1$)
\begin{eqnarray}
\Gamma_0 (\mbox{O-Ps}
\rightarrow 3 \gamma)&=& \vert \Psi(0)\vert^2 ~\hat \sigma_0 (0)
  \nonumber
 \\ &=&  { 2 \pi^2 -18 \over 9 \pi} \, m_e \alpha^6
 \nonumber \\ &=& 7.2112 \, \mu {\rm s }^{-1}.   \label{rate}
\end{eqnarray}
Here $\Psi(0)$ is the ground state Schr\"odinger-Coulomb
wavefunction evaluated at $\vec r =0$, and $\hat \sigma_0 (p)
$ is proportional to the lowest-order decay rate of a {\em free} electron and
positron in an S-state:
\begin{eqnarray}
\hat \sigma_0 (p) \equiv  {1 \over 4 m_e^2}
   ~{\rm Im}~ {\cal M}_0^{(l=0)}\, \bigl( e
\overline{e}   \rightarrow 3 \gamma \rightarrow e \overline{e} \bigr) ,
\end{eqnarray}
where $p$ is the magnitude of the electron center-of-mass momentum.

 To order ${\cal O } (\alpha^2)$, there are three sources of corrections:

\medskip
\noindent   {\sc 1)    RADIATIVE CORRECTIONS TO $\hat \sigma_0 (0)$:}
 These corrections
renormalize the lowest-order decay rate:
\begin{eqnarray}
\delta \Gamma_1 ~=~ \delta Z ~ \Gamma_0 (\mbox{O-Ps} \rightarrow 3 \gamma) .
\label{aa}
\end{eqnarray}
The ``renormalization'' constant $\delta Z$ is defined to be  the part of the
radiative corrections coming from relativistic loop momenta. Thus it
is insensitive to the binding energy of the atom, and can be related
to the infrared-finite part of the radiatively corrected annihilation
rate $\hat \sigma$ for a
  free electron
and positron at $p=0$:
\begin{eqnarray}
\hat \sigma (0) - \hat \sigma_{{\rm IR}} (0) ~=~ ( 1 + \delta Z) \, \hat
\sigma_0(0) , \label{ddd}
\end{eqnarray}
where $ \hat \sigma_{{\rm IR}} (0)$ is the infrared part of $\hat \sigma (0)
$. Using a photon mass to regulate the infrared (threshold)
 singularities\footnote{The precise definition of  $\hat \sigma_{{\rm IR}} (0)$
 depends upon the regulator used in the NRQED part of the analysis; see
refs \cite{NRQED}, \cite{hfs}. }
\begin{eqnarray}
{ \hat \sigma_{{\rm IR}} (0) \over \hat
\sigma_0(0)} &   =& ~ { m_e \alpha \over \lambda} +
  ( 2 \ln 2 + 1 ) \alpha^2  \, { m_e^2 \over \lambda^2} \nonumber \\
&  -&  2 \times 10.282(3) {\alpha^2 \over \pi} { m_e \over
\lambda} + {1 \over 3} \alpha^2 \, \ln {\lambda \over m_e} +{\cal O}(\alpha^3)
 . \label{dd}
\end{eqnarray}
 These infrared  subtractions
remove all contributions due to nonrelativistic loop momenta. Thus
$\delta Z$ has a $\lambda$-independent expansion in powers of $\alpha/\pi$:
\begin{eqnarray}
\delta Z~=~ c_1 { \alpha \over \pi} + c_2 \left( {\alpha \over \pi} \right)^2
 + \ldots  . \label{ee}
\end{eqnarray}
The only $ {\cal O} (\alpha)$ corrections to the decay  rate come from $\delta
Z$,  and  thus,  from Eq.[\ref{expansion}]
$c_1 = -10.282(3)$. The second-order coefficient has not yet been calculated.
The large size of $c_1$ suggests that $c_2$ might also be large. Indeed,
a small (gauge invariant) subset of the diagrams contributing to
$c_2$ has recently been evaluated and found to
contribute $28.8(2)$ to $c_2$ \cite{Russia}.

\medskip
\noindent  {\sc 2) MOMENTUM DEPENDENCE OF  $\hat \sigma_0 (p)$ :} Near
threshold, the decay rate for a free electron and  positron in an
S-wave has the form
\begin{eqnarray}
\hat \sigma_0 (p) ~=~ \hat \sigma_0 (0) + \delta \hat \sigma_0 (p)
+ {\cal O} ( {  p^4  \over m_e^4} \hat \sigma_0  ) , \label{cc}
\end{eqnarray}
where $\delta \hat \sigma_0 (p) $ can be computed to be \cite{papier}:
\begin{eqnarray}
\delta \hat \sigma_0 (p) ~=~-
  {  p^2 \over m_e^2} ~  {  19 \pi^2
- 132 \over 12 ( \pi^2 - 9) }~   \hat \sigma_0 (0)  .
\end{eqnarray}
This correction term shifts the decay rate in ${\cal O} (\alpha^2)$.
 Na\"\i vely, one might expect a contribution
\begin{eqnarray}
 2 \Psi(0) \, \int { d^3 p \over (2 \pi)^3}
\, \delta \hat \sigma_0 (p) \, \Psi( p) . \label{bb}
\end{eqnarray}
However, this integral includes contributions already present in the
relativistic radiative corrections discussed above (Eq.[\ref{aa}]).
 This becomes
 apparent if the Schr\"odinger equation is used to rewrite the wavefunction:
\begin{equation}
\Psi( p)~=~ { 1 \over E_{\mbox{O-Ps}} -  p^2/m_e}~\int { d^3 q \over
 (2 \pi)^3} \, V_C ({\bf p} - {\bf q})\,  \Psi( q)
\end{equation}
where $V_C$ is the Coulomb potential, $V_C({\bf p} - {\bf q}) = -e^2/({\bf p} -
{\bf q} )^2 $, and $E_{\mbox{O-Ps}} = - m \alpha^2/4 $ is the binding energy.
Then our contribution to the orthopositronium decay rate becomes
\begin{equation}
2\,  \Psi(0) \, \int {d^3 p \over (2 \pi)^3} \,
 \, \delta \hat \sigma_0 (p)\,  \Psi( p)
  ~=~2\,  \Psi(0) \int {d^3p \over (2 \pi)^3}\,
  \delta \hat \sigma_C(E_{\mbox{O-Ps}},
  p)
\, \Psi( p)
\end{equation}
where
\begin{equation}
\delta \hat \sigma_C(E_{\mbox{O-Ps}},
  p)
  ~\equiv~
 \int {d^3q \over (2 \pi)^3}~  \delta \hat \sigma_0 ( \vert {\bf p}
+ {\bf q} \vert  )~ {1 \over E_{\mbox{O-Ps}} - ({\bf p} + {\bf q})^2/m_e}
  ~  V_C ( q)
\end{equation}
is a one-loop radiative correction to the basic decay.

To avoid
double-counting the relativistic radiative corrections, we subtract
 the threshold value of
$\delta \hat \sigma_C $, {\it i.e.}
\begin{equation}
\delta \hat \sigma_C (E,  p)  \rightarrow \delta \hat \sigma_C
 (E,  p) - \delta \hat \sigma_C (0,0).
\end{equation}
Thus the final contribution to the decay rate from $\delta \hat \sigma_0 (p)$
is
\begin{equation}
\delta \Gamma_2 ~=~ 2 \Psi(0) \int {d^3 p \over (2 \pi)^3} \left( \delta
\hat \sigma_C (E_{\mbox{O-Ps}},  p) -
\delta \hat \sigma_C (0,0) \right) \Psi ( p) .
\end{equation}
Since $\delta \hat \sigma_0 (p) \propto  p^2$, this result simplifies to
\begin{eqnarray}
\delta \Gamma_2 ~& =& -  { {\rm E_{\mbox{O-Ps}}} \over m_e} ~ { 19 \pi^2
- 132  \over  12 (  \pi^2 -9 )  }  \, \hat \sigma_0 (0) \nonumber
\\ & \approx &
5 \times 10^{-4} \mu {\rm s}^{-1} .
\end{eqnarray}

Note that our original expression for this contribution
(Eq.[\ref{bb}])
 is ultraviolet divergent. The divergence comes from the high-$ p$
region, which is precisely the region that must be removed
so as to avoid double counting the relativistic radiative correction.
Consequently, our subtraction regulates the integral. Such infinities
are common when nonrelativistic expansions like that for $\hat \sigma_0 (p)$
 (Eq.[\ref{cc}]) are employed. However, they can always
be removed by appropriate counterterms. The structure of these
counterterms is readily determined using NRQED. The
use of this theory puts calculations such as ours on a
rigorous foundation, permitting  systematic improvement
of the final results (a  rigorous derivation  of the NRQED expansion to
second order in $ p /m$ and $\alpha$,
including all  counterterms,  will be  derived
 in another paper
\cite{hfs}).

\medskip
\noindent  {\sc 3) CORRECTIONS TO $\Psi (0) $ :} The wavefunction at the origin
in the lowest-order rate (Eq.[\ref{rate}]) is modified by the various
relativistic corrections that must be added to the
Coulomb-Schr\"odinger theory. Using first-order perturbation theory,
the correction is
\begin{eqnarray}
\delta \Psi(0)~=~ \sum_{n \neq \mbox{O-Ps}} \Psi_n(0) { \langle n | \, \delta
 V \, | \mbox{O-Ps} \rangle
\over E_{\mbox{O-Ps}} - E_n }
\end{eqnarray}
where the sum is over all eigenstates of the unperturbed theory, and
\begin{eqnarray}
\delta V({\bf p} ,
 {\bf q})~& = &~ - (2 \pi)^3 \delta ({\bf p} - {\bf q}) \, {  p^4
\over 4 m_e^3 }  - { e^2 \over m_e^2  q^2 }\,  \bigl(  p^2
- { ({\bf p} \cdot {\bf q})^2 \over  q^2} \bigr)  \nonumber
\\  & & +{ 3 e^2 \over 4 m_e^2
}   - { e^2 \over 4 m_e^2}\,
\mbox{\boldmath $\sigma$}_1 \cdot \mbox{\boldmath $\sigma$}_2
 + { e^2 \over 4 m_e^2 } {({\bf q} \cdot
 \mbox{\boldmath $\sigma$}_1 ) \, ( {\bf q} \cdot \mbox{\boldmath $\sigma$}_2
 ) \over q^2 }
\nonumber \\ && - 3 { \alpha^2 \over m_e^2} \ln( q^2/m_e^2) .
\end{eqnarray}
The potential $\delta V$ contains all ${\cal O} (v^2/c^2)$ corrections to the
Schr\"odinger theory as well as the leading ${\cal O} (\alpha \, v^2/c^2)$
correction, given by the $\ln ( q^2/m_e^2)$ term. This last term
gives the leading contribution to the Lamb shift in positronium.

The corrections to the wavefunction will  shift the decay rate by
\begin{eqnarray}
\delta \Gamma_3 ( \mbox{O-Ps}
 \rightarrow 3 \gamma) ~=~ 2 \, \vert \delta \Psi(0)
\, \Psi(0) \vert~ \hat \sigma_0 (p) .
\end{eqnarray}
As in the case of $\delta \hat \sigma (p)$,  the ultraviolet divergences
 encountered in evaluating $ \delta \Psi(0) $
are systematically  removed by using NRQED.
The remaining contributions all come from nonrelativistic loop
momenta. We find
\begin{eqnarray}
\delta \Psi(0) ~=~ \bigl[ { 1 \over 6} \alpha^2 \ln \alpha + 0.58 \alpha^2
 - { 3 \over 4  \pi} \alpha^3 (\ln \alpha)^2 \bigr]~\Psi(0)
\end{eqnarray}
which implies
\begin{eqnarray}
\delta \Gamma_3 ~=~\bigl[ { 1 \over 3} \alpha^2 \ln \alpha + 1.16\, \alpha^2
 - { 3 \over 2 \pi} \alpha^3 (\ln \alpha)^2
 \bigr]~ \Gamma_0 ( \mbox{O-Ps} \rightarrow 3 \gamma) .
\end{eqnarray}
The logarithmic terms agree with  the literature \cite{Lepage,Karsh}; the
other correction is new.

By combining  the ${\cal O} ( \alpha ^2 \Gamma_0)$  contributions
 from
all three of our  corrections, we obtain the final result
\begin{eqnarray}
\delta \Gamma_1 + \delta \Gamma_2 + \delta \Gamma_3 \vert_{\alpha^2}
  & =& \biggl\{
{ 19 \pi^2
- 132  \over 12 (  \pi^2  -9 )  } \, { \alpha ^2 \over 4 }
+  1.16 \, \alpha^2
+ c_2 \left( {\alpha \over \pi} \right)^2 \biggr\} \Gamma_0  \nonumber \\
& = &
9.56 \times 10^{-4}
  \, \mu {\rm s}^{-1} \, +  c_2 ~ 0.39 \times 10^{-4} \, \mu {\rm s}^{-1}  .
\end{eqnarray}
The coefficient $c_2$ is specified by the threshold rate for $e \overline{e}
 \rightarrow
3 \gamma$ (Eqs.[\ref{ddd}, \ref{dd}, \ref{ee}]).
The known part of $c_2$ \cite{Russia}
 contributes $11 \times 10^{-4} \mu {\rm s}^{-1} $
to the rate.

In this paper, we have outlined a new procedure for analyzing the \mbox{O-Ps}
 $\,$
 decay
rate at ${\cal O} (\alpha^2 \Gamma_0) $.
 We have computed all corrections in this
order that depend in detail on bound-state physics. The only remaining
contribution can be extracted from a calculation of the annihilation rate
for a free electron and positron -- no bound-state physics  is required.
Our corrections combined with the known part of the $c_2$ correction account
for 20\% of the difference between theory and experiment. The unexpectedly
large size of these corrections makes it plausible  that the full
calculation of $c_2$ will bring theory into
agreement with experiment.

 U.M. thanks the Academy of Finland for support and Cornell for hospitality
during the completion of this work.
P.L. has been supported by a ``1967" NSERC fellowship (Canada)
  and by  the  Fonds Canadiens \`a la recherche (Qu\'ebec).
This work was also supported by a grant from the National Science
Foundation.
We thank Todd Ward for helpful comments concerning the analytical calculation
of the $ p^2/m_e^2$ correction.
\vfill
\eject

\end{document}